\documentclass[]{raa}
\usepackage{graphicx}
\usepackage{times}
\usepackage{natbib}
\usepackage{mathrsfs}
\usepackage[justification=centering]{caption}
\usepackage{amsmath}
\usepackage{amssymb}
\usepackage{booktabs}
\usepackage{hyperref}
\usepackage{color}
\usepackage{multirow}
\usepackage{threeparttable}
\hypersetup{colorlinks = true, linkcolor = green, anchorcolor = red, citecolor = blue, filecolor = red, pagecolor = red, urlcolor = red}

\begin{document}

   \title{Revise thermal winds of remnant neutron stars in gamma-ray bursts}
   \volnopage{Vol. 000 No.0, 000--000}      
   \setcounter{page}{1}          
   \author{
    Shuang Du
      \inst{1}
    \and Tingting Lin
    \inst{2}
   \and Shujin Hou
      \inst{3}
   \and Renxin Xu
      \inst{4,5}
   }

    \institute{%
           School of Mathematics and Computer, Tongling University, Tongling 244061, Anhui, China; {\em dushuang@pku.edu.cn} \\
      \and
           School of Foundation, Zhejiang Shuren University, Shaoxing 312028, Zhejiang, China  \\
      \and
           School of Physics and Electronic Engineering, Nanyang Normal University, Nanyang, Henan 473061, China
       \and
           School of Physics, Peking University, Beijing 100871, China; {\em r.x.xu@pku.edu.cn} \\
        \and
           Kavli Institute for Astronomy and Astrophysics, Peking University, Beijing 100871, China
   }

\date{Received~~2023 xx xx; accepted~~20xx xx xx}

\abstract{
It seems that the wealth of information revealed by the multi-messenger observations of the binary neutron star (NS) merger event, GW170817/GRB 170817A/kilonova AT2017gfo, places irreconcilable constraints to models of the prompt emission of this gamma-ray burst (GRB).
The observed time delay between the merger of the two NSs and the trigger of the GRB and the thermal tail of the prompt emission can hardly be reproduced by these models simultaneously.
We argue that the merger remnant should be an NS (last for, at least, a large fraction of $1\rm s$), and that the difficulty can be alleviated by the delayed formation of the accretion disk due to the absorption of high-energy neutrinos emitted by the NS and the delayed emergence of an effective viscous in the disk.
Further, we extend the consideration of the effect of the energy deposition of neutrinos emitted from the NS.
If the NS is the central object of a GRB with a distance and duration similar to that of GRB 170817A,
thermal emission of the thermal bubble inflated by the NS after the termination of accretion may be detectable.
If our scenario is verified, it would be of interest to investigate the cooling of nascent NSs.
 }
  \maketitle

\section{Introduction}\label{sec1}

The multi-messenger observations of the binary neutron star (NS) merger event, GW170817/GRB 170817A/kilonova AT2017gfo,
provide abundant information for relevant topics \citep{2017ApJ...848L..12A}, for example, the origin of short gamma-ray bursts
(sGRBs; e.g., \citealt{2017ApJ...848L..13A}; see \citealt{2021ARA&A..59..155M} for a review).
However, the central object left by the binary NS merger is still uncertain \citep{2017ApJ...851L..16A,2019ApJ...875..160A}.
Some works have discussed the connection between the merger remnant
and the corresponding kilonova \citep{2017ApJ...850L..19M,2018ApJ...861..114Y} and the GRB afterglow \citep{2019MNRAS.483.1912P,2022ApJ...927L..17H,2022MNRAS.512.5572R}.
Likewise, there may be also a certain connection between the remnant and the GRB prompt emission,  
for example, \cite{TYH} had considered a magnetar collapse scenario for the prompt emission (but, this model is not supported by the afterglow observation and difficult to explain the spectral evolution of the prompt emission).

The burst triggered $\sim 1.74\rm s$ after the merger of the two NSs \citep{2017ApJ...848L..13A}.
\emph{Fermi}-GBM observation of GRB 170817A \citep{2017ApJ...848L..14G}
shows that the burst has two periods that a main pulse from $t_{\rm t}-0.320\rm s$ to $t_{\rm t}+0.256\rm s$
and a lower-significance tail from $t_{\rm t}+0.832\rm s$ to $t_{\rm t}+1.984\rm s$, where $t_{\rm t}$ is the GRB trigger time.
The spectrum of the main pulse shows a non-thermal feature that can be best fitted by a Comptonized function.
However, the spectrum of the weak tail indicates a black-body radiation with temperature $T_{\rm th}=10.3\pm 1.5\rm keV$.

The abundant observations give many constraints on GRB models and make these models faultily (see \citealt{2021ARA&A..59..155M} for a review).
The basic difficulty is that, if the thermal tail of GRB 170817A is reliable,
physical models are hard to naturally reproduce the observed time delay of $\sim 1.74\rm s$ and the black-body radiation with temperature $T_{\rm th}=10.3\pm 1.5\rm keV$, simultaneously.

In Section \ref{sec2}, we clarify the conflictions between observations and models and further show a solution to the difficulty that, if the merger remnant is an NS,
the delay can be partly resulted by the delayed disk formation due to the absorption of high-energy neutrinos emitted by the NS and delayed emergence of effective viscous in the accretion disk.
In Section \ref{sec.3}, we extend the consideration shown in Section \ref{sec2} and discuss the electromagnetic signal of the thermal bubble inflated by the NS.
Section \ref{sec4} is the summary and discussion.

\section{conflictions and the solution}\label{sec2}
\subsection{Conflictions}\label{sec2.1}
The prompt emission of GRB 170817A is different from previous GRBs with thermal components being detected.
There is no such an evident interval and boundary between the power-law and thermal components in pervious ones (as summarized in \citealt{2019ApJS..245....7L}).
As shown in Figure 6 of \cite{2017ApJ...848L..14G}, the flux of the soft tail is compatible to that of the main pulse in $10-50\rm keV$.
This means that if there is a similar thermal component in the main pulse, this thermal component should be visible too \footnote{This is an important aspect that has been overlooked and indicates the separation between the radiation region of the main gamma-ray pulse and the soft thermal tail.}.
The Comptonized component followed by a thermal component challenges the classical framework of GRBs.
(i) According to the widely accepted model that the prompt emission of a GRB is powered by the energy release in the GRB jet after the jet turning into optically thin (e.g., \citealt{1994ApJ...430L..93R}; see \citealt{2018pgrb.book.....Z} for more details),
if the jet is hot enough at the photosphere, the thermal component should be detected at the first but not following the Comptonized main pulse (see also in \citealt{2017ApJ...848L..13A});
(ii) No matter how the jet is produced, the jet should be uniform on its component at the line of sight (but not the stratified jet; \citealt{2021RAA....21..177C}).
The production of the jet that a cold segment followed by a hot segment is confusing.
All these facts indicate that just considering the dissipation of jet energy (e.g., \citealt{1994ApJ...430L..93R,2011ApJ...726...90Z}) cannot explain the time delay and thermal tail, uniformly. To explaining these characteristics, some geometrical effects should be invoked.

The absence of evident thermal component in the main pulse implies that there should be a separation between the radiation regions (as well as relevant parameters) of the two components.
\cite{2018ApJ...860...72M} showed that an off-axis high-latitude-enhanced photosphere emission plus an on-axis photosphere emission of a structured jet can explain the observed spectrum.
However, the Lorentz factor along the line of sight of the jet is too large, $\sim 20$, which is deviated from the constraint of the GRB afterglow (e.g., see Figure 7 in \citealt{2021ARA&A..59..155M}),
and a significant delay between the merger and jet launching is required since the photosphere is too close to the central engine so that the time delay of $1.74\rm s$ cannot be totally
attributed to the time wasting of jet moving. \cite{2018MNRAS.479..588G} showed another scenario that the hard-to-soft spectral evolution can be induced by the transition (a planar to a spherical phase) of the emission arose from a shock breakout of a cocoon from the merger's ejecta. Similarly, this scenario also requires a delayed jet injection.
Therefore, to explain the time delay between GRB 170817A and GW170817 and the hard-to-soft spectral evolution simultaneously,
under both of the structured-jet scenario and jet-cocoon scenario (should be the only two jet profiles that allowed by the observed properties of GRB 170817A; \citealt{2017ApJ...848L..13A}; see \citealt{2019ApJ...877L..40G} for simulation side),
a delayed jet injection is required.
\cite{2017ApJ...850L..24G} suggested that the merger remnant is a short-lived (lifetime $\leq 1 \rm s$) hypermassive NS, and the jet only can be launched after the hypermassive NS collapses into a black hole. However, this consideration is difficult to be supported by both of theory (see, \citealt{2017NewAR..79....1L} for a review; even the remnant is an NS, the accretion is still ongoing) and observation \citep{2022ApJ...939..106J}.

\subsection{A possible solution}
Under the classical GRB framework (see \citealt{2018pgrb.book.....Z} for more details),
the delayed jet injection indicates the delayed disk formation since the jet is launched by the central object-disk system.
Simulations show that the disk is formed during several dynamical time scale of the system after the merger ($t_{\rm for}\sim 10{\rm ms}$) regardless of the central object \citep{2019ARNPS..69...41S}.
However, the type of the central object may affect the formation of the disk.
According to the observations of SN1987A \citep{1987IAUC.4316....1K,1987PhRvL..58.1490H,1987PhRvL..58.1494B},
the temperature of the nascent NS is several $\rm MeV$ and the NS can stay very hot for a few seconds after the birth (see \citealt{1989ARA&A..27..629A} for a review).
If the central object of GRB 170817A is an NS, comparing with the black-hole case, there will be an extra energy shedding from the central engine that
a strong neutrino emission and neutrino-driven wind can emerge from the nascent NS \citep{1986ApJ...309..141D,1996ApJ...471..331Q}.
The absorption of high-energy neutrinos may turn the gravity-bound ejecta into free
so that the formation of the disk will be put off until some of the ejecta falls back due to, for example, viscosity and friction in the ejecta and the cooling of the NS.
Therefore, the disk formation would be determined by the competition between the release of gravitational potential energy and energy deposition in the gravity-bound ejecta.
The time necessary for the gravity-bound ejecta to absorb enough energy to overcome the gravity can be estimated as \citep{2014MNRAS.443.3134P}
\begin{eqnarray}\label{eq1}
t_{\rm fre}\sim\frac{e_{\rm gra}}{\dot{e}_{\rm neu}}= 0.02{\rm s}\left ( \frac{M_{\rm ns}}{2.5{\rm M_{\odot}}} \right )\left ( \frac{R_{\rm in}}{30\rm km} \right )\left ( \frac{L_{\nu_{\rm e}}}{3\times 10^{52}\rm erg\;s^{-1}} \right )^{-1}\left ( \frac{\xi}{1.5}  \right )^{-1}\left ( \frac{E_{\nu_{\rm } }}{15\rm MeV} \right )^{-2},
\end{eqnarray}
where $e_{\rm gra}$ is the specific gravitational energy, $R_{\rm in}$ is the inner radius of the disk which is assumed to be close to the
innermost stable circular orbit of a Schwarzschild black hole with the same mass as the NS, $\dot{e}_{\rm neu}$ is the specific heating rate provided by neutrino absorption at a
radial distance $R_{\rm in}$ from the centre, $M_{\rm ns}$ is the NS mass, $L_{\nu_{\rm e}}$ is the luminosity of electron neutrinos from the NS,
$\xi$ is the factor to describe the aspect ratio of the disk, and $E_{\nu_{\rm }}$ is the neutrino energy from the NS.
Comparing $t_{\rm for}$ and $t_{\rm fre}$, one can find that they are of the same order, that is, it is possible to put off the formation of the disk through the absorption of high-energy neutrinos\footnote{If the value of $t_{\rm fre}$ is smaller than that of $t_{\rm for}$ but larger than the duration of the accretion, there will be a pause of the accretion, as well as the jet launching, when the accretion last for $t_{\rm fre}$, and then energy release in the foregoing jet will appear as a precursor. }.
The upper limit of the delay of disk formation due to the absorbing of neutrinos can be estimated as the cooling timescale of the NS, that is \citep{2014MNRAS.443.3134P}
\begin{eqnarray}
t_{\rm cool,ns}\sim \frac{3R_{\rm ns}^{2}}{c\lambda_{\rm N\nu }}\approx  0.7{\rm s}
\left ( \frac{R_{\rm ns}}{15\rm km}  \right )^{2}\left ( \frac{\rho_{\rm ns}}{10^{14}\rm g\; cm^{-3}}  \right )
\left ( \frac{E_{\nu_{\rm } }}{15\rm MeV} \right )^{2},
\end{eqnarray}
where $R_{\rm ns}$ is the radius of the NS, $\lambda_{\rm N\nu }$ is the mean free path of neutrinos in the NS which is given by
\begin{eqnarray}
\lambda_{\rm N\nu }\approx 7.44\times 10^{3}{\rm cm}
\left ( \frac{\rho_{\rm ns}}{10^{14}\rm g\; cm^{-3}}  \right )^{-1}
\left ( \frac{E_{\nu_{\rm } }}{15\rm MeV} \right )^{-2},
\end{eqnarray}
and $\rho_{\rm ns}$ is the matter density of the NS.

Besides, to start the accretion and launch the jet (e.g., through neutrino pair annihilation; see \citealt{2017NewAR..79....1L} for a review)\footnote{The tangible surface of the NS will prevent Blandford-Znajek process \citep{1977MNRAS.179..433B} even if the NS has an ergosphere. However, other collimation mechanisms for Poynting-flux-dominated jets under magnetar case may exist (e.g., \citealt{2012MNRAS.419.1537B}).}, the disk should transfer angular momentum and generate thermal energy.
It is difficult to estimate when the viscous works. If only the main gamma-ray pulse of GRB 170817A is produced by the jet and the thermal tail originates from other reasons (e.g., from the cocoon), to match the duration of the main pulse $t_{\rm 90,m}\approx 0.58\rm s$ \citep{2017ApJ...848L..14G},
the viscosity parameter \citep{1973A&A....24..337S} can be estimated as
\begin{eqnarray}
\alpha\sim t_{\rm 90,m}^{-1}\left ( \frac{H}{R_{\rm out}} \right )^{-2}\Omega_{\rm k}^{-1}\approx 0.02 \left ( \frac{t_{\rm 90,m}}{0.6\rm s}  \right )^{-1}
\left ( \frac{H/R}{1/3} \right )^{-2}\left ( \frac{M_{\rm ns}}{2.5\rm M_{\odot}} \right )^{-1/2}\left ( \frac{R_{\rm out}}{100\rm km} \right )^{3/2},
\end{eqnarray}
where $H$ is the disk height, $H/R$ is the aspect ratio of the disk, and $R_{\rm out}$ is the outmost radius of the disk.
Usually, without some magnetic instabilities, the value of the viscosity parameter cannot be up to $\sim 0.01-0.1$ (see \citealt{2008bhad.book.....K} for more details).
Therefore, if effective magnetic field cannot emerge in the disk timely, the jet launching may be also put off.
As an example, if the seed of the magnetic field in the disk is from the NS, only the magnetic field of the NS is generated (e.g., in $\sim 1\rm s$ after the formation of the proto-NS; \citealt{1993ApJ...408..194T}), can the disk begin to amplify this magnetic seed.
Note that, a large-scale magnetic field in the disk should be essential. The magnetic field can prevent proton deposition in the jet core,
so that excessive mass loading due to the neutrino-driven wind in the jet \citep{2014MNRAS.443.3134P} will be avoided.

\subsection{An interim summary}

Based on the assumptions that the thermal gamma-ray tail of GRB 170817A is real and
the structured jet and the jet-cocoon structure are the only two allowed jet profiles to explain the thermal tail, we get the following claims.
(a) The incompatibility between observations and models of GRB 170817A indicates the requirement of the delayed jet injection.
(b) Although quantitative estimation is absence due to the unknown cooling and magnetization of nascent NSs,
neutrino deposition and delayed emergence of effective viscous in the disk should be physical mechanisms that may result a fraction of the time delay of $\sim 1.74\rm s$
(e.g., $\sim 0.7\rm s$ after the merger as required in \citealt{2018MNRAS.479..588G}).

The delay may not be the only effect of the energy deposition of neutrinos in GRBs. As long as the duration of a GRB is short enough and the central object of the GRB is an NS,
when the accretion terminates, the NS could be still hot enough,
so that a clearer fireball (compared with the neutrino-driven wind launched before the termination of the accretion) will be inflated through neutrino pair annihilation \citep{1987ApJ...314L...7G} and even thermal radiation \citep{2001ApJ...550L.179U}.
Although, GRBs should be powered by energy release in jets, it is meaningful to revise the evolution of the fireball since it is directly powered by the NS and may reveal the information of the GRB central engine and test above idea. In the next section, we discuss possible electromagnetic counterpart of the fireball.

\section{Possible connection between NSs and precursors}\label{sec.3}
\subsection{Basic schematic diagram}\label{sec3.1}
\begin{figure}
\centering
\includegraphics[width=0.6\textwidth]{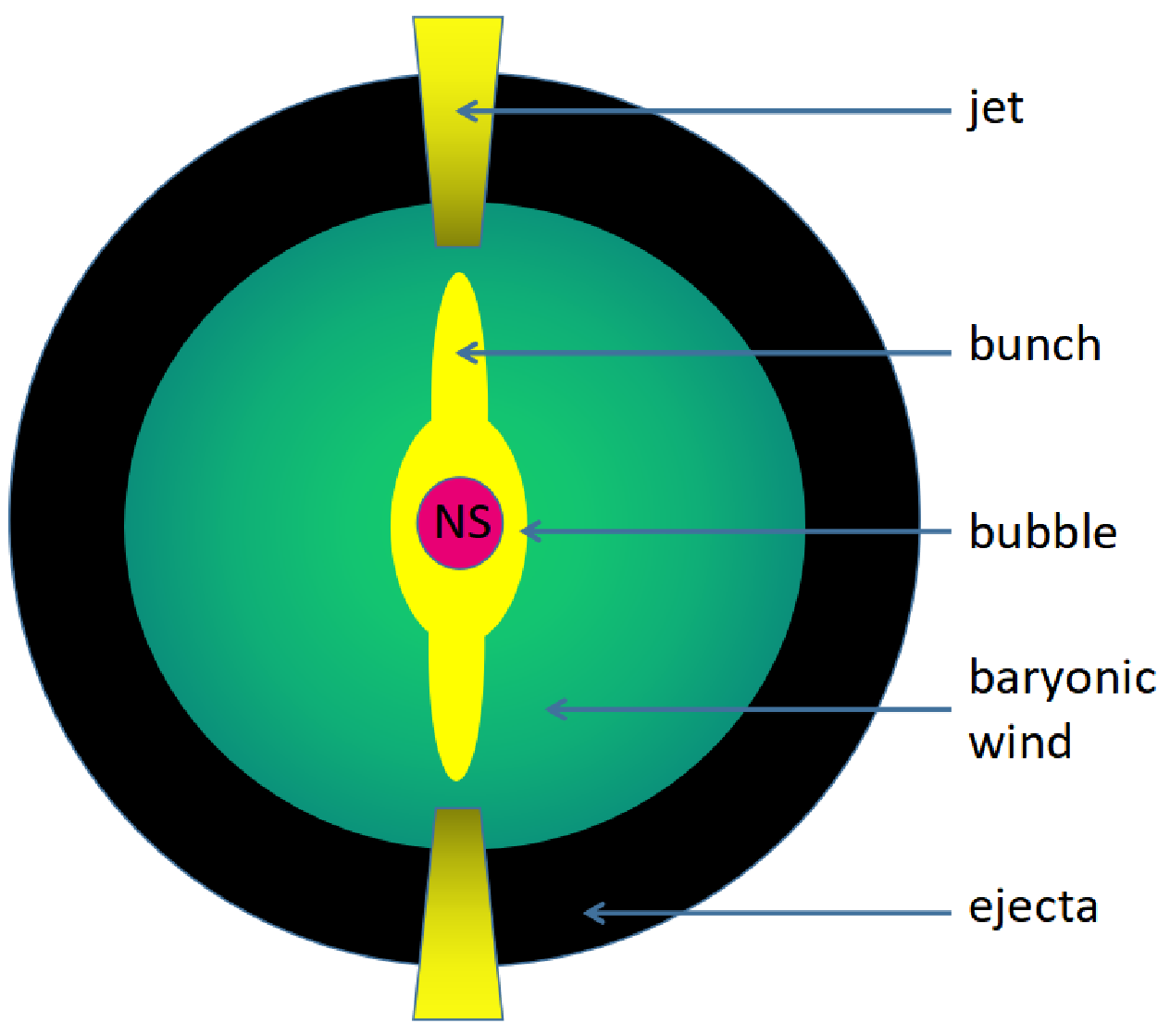}
\caption{Schematic diagram for the evolution of the thermal bubble. }
\label{fig1}
\end{figure}

After the onset of the binary NS merger, the dynamically mass ejection begins to be ejected, and the proto-NS soon forms later, as well as the neutrino-driven wind from the NS \citep{1986ApJ...309..141D,1996ApJ...471..331Q}\footnote{Note that the neutrino-driven wind will also impact the gravity-bound ejecta and impede the accretion. However, once the system begins to accrete matter, it means that the release of gravitational potential energy dominates the system and the energy deposition due to absorbing neutrinos becomes a secondary effect. }.
Almost simultaneously (e.g., in $\sim 1\rm s$ after the merger according to the above discussion), an accretion disk emerges.
The disk and the NS make up the GRB central engine, so that the relativistic jet begins to be launched.
Together with the jet launching, baryonic wind driven by neutrinos from the NS and disk \citep{1997A&A...319..122R,2009ApJ...690.1681D,2014MNRAS.443.3134P} is accompanying.
After the termination of the accretion,
a fireball with less baryonic mass loading and totally powered by the NS is inflated.
As the cooling of the NS is very fast \citep{1986ApJ...307..178B}, this fireball should be less energetic than the fireball expected in \cite{1986ApJ...308L..43P,1990ApJ...363..218P}.
Hereafter, this less energetic and clearer fireball is called as thermal bubble.
Thanks to the jet breaking the ejecta, the interspace left by the jet will be continually filled up by the matter extended from the thermal bubble.
That is, the thermal bubble will mainly inflates along the direction of the jet moving, and the inflation along other directions will be suppressed by this prior distribution of energy and the constraint of surrounding matter.
Correspondingly, two bunches will extend from the thermal bubble (see Fig. \ref{fig1}).
As long as the power of the bunch is large enough,
the bunch will rush out from the ejecta following the jet ultimately.
Empirically, according to the isotropic luminosity of on-axis low-luminosity GRBs (e.g., GRB 060218, \citealt{2006Natur.442.1008C}; GRB 100316D, \citealt{2011MNRAS.411.2792S})\footnote{The fluxes of the X-ray afterglows of these GRBs are continuous attenuation which is different from that of the off-axis GRB (GRB 170817) and consistent with the expectation of on-axis GRBs. Therefore, we believe these GRBs are viewed on-axis and cannot be explained via jet-cocoon scenario.},
the critical isotropic luminosity of the bunch which can make the bunch break out from the surrounding matter should be
\begin{eqnarray}\label{eq2}
L_{\rm c}\sim 10^{46}{\rm erg\; s^{-1}}.
\end{eqnarray}
Although the cooling of the nascent NS is highly uncertain \citep{2004ARA&A..42..169Y,2006NuPhA.777..497P},
observation \citep{1989ARA&A..27..629A} and simulation \citep{1996ApJ...471..331Q} indicate that
the temperature of the nascent NS should maintain $10^{10}-10^{9}\rm K$ for several seconds.
Therefore, as long as the lifetime of the accretion disk is short enough, the total energy of the thermal bubble inflated in several seconds may be up to $10^{48}-10^{49}\rm erg$
as expected in \cite{1987ApJ...314L...7G}.
So, the luminosity of the bunch can beyond the critical value.
With the cooling of the NS, the luminosity of the thermal bunch will be larger than the power supplied by the NS at a certain time.
If the bunch gets out of the encirclement after this time point, the totally energy of the thermal bubble will be reduced,
as well as the energy density and pressure.
Therefore, ultimately, the escape route left by the jet will be closed by the squeezing of the surrounding matter due to the decrease of the energy density in the thermal bubble,
and then the bunch will be cut off. After the breakout, the bunch will be finally transparent for the thermal radiation.
In the next, we discuss the detectability of the thermal radiation.

\subsection{The photosphere emission of the bunch}\label{sec3.2}
If the bunch can break out of the ejecta, the fundamental problem is when or where the breakout occurs.
We need to know the velocities of the ejecta and bunch.
The expansion time of the ejecta is $\sim R_{\rm ns}/v_{\rm e}$, where $R_{\rm ns}$ is the radius of the NS, and $v_{\rm e}$ is the velocity of the ejecta.
According to the duration of sGRBs, the lifetime of the accretion disk should be more likely longer than $R_{\rm ns}/v_{\rm e}$.
So, after the termination of the accretion, the ejecta should be in the self-similar adiabatic expansion stage with an constant velocity ($\sim 0.1 c-0.3 c$, where $c$ is the speed of light).
Similarly, the bunch also soon inflates to the maximum speed since the faster jet has cleared the way for the bunch\footnote{If the jet power is sufficiently larger than the critical value defined by equation (1), we can expect that the acceleration of the jet is almostly unaffected by the ejecta. Then the acceleration of the followed bunch is also unaffected since the jet should be faster.}.
The saturation radius of the bunch can be estimated as $R_{\rm b,c}\sim \eta_{\rm b}R_{\rm ns}$ \citep{1993MNRAS.263..861P}, where $\eta_{\rm b}$ is the dimensionless entropy of the bunch
(i.e., the maximum Lorentz factor of the bunch).
The Lorentz factor of the jet is usually $\eta_{\rm j}\sim 10^{2}-10^{3}$. Comparing with the jet, the bunch is less energetic and clearer.
Therefore, the order of magnitude of the maximum Lorentz factor of the bunch that $\eta_{\rm b}\sim 10^{1}$ is adopted,
and then the saturation radius of the bunch is $R_{\rm b,c}\simeq \eta_{\rm b}R_{\rm ns}\sim 10^{7}\rm cm$.
If the breakout radius of the thermal bunch, $R_{\rm b,b}$, is large enough that $R_{\rm b,b}\gg R_{\rm b,c}$,
the velocity of the bunch is relativistic in the most of the time before breaking out of the ejecta.
Under this case, the breakout radius can be estimated as
\begin{eqnarray}\label{eq110}
R_{\rm b,b}&\sim &\frac{{v}_{\rm b}v_{\rm e}\Delta t}{{v}_{\rm b}-v_{\rm e}}=3.0\times 10^{9}{\rm cm}\left ( \frac{v_{\rm e}}{0.1c}\right )\left ( \frac{t_{\rm a}}{1\rm s} \right ) \left (\frac{{v}_{\rm b}/({v}_{\rm b}-v_{\rm e})}{1} \right ),
\end{eqnarray}
where $\Delta t$ is the time interval between the ejection of the ejecta and the inflation of the thermal bubble,
and ${v}_{\rm b}\approx c$ is the velocity of the bunch.
Comparing $R_{\rm b,c}$ and $R_{\rm b,b}$, one can find that the estimation is self-consistent.

To see the thermal radiation from the bunch,
the bunch must reach its photosphere radius, $R_{\rm ph,b}$, and there should be $R_{\rm ph,b}>R_{\rm b,b}$.
As discussed above, the saturation radius of the bunch is much smaller than the breakout radius of the bunch.
Therefore, only the case that $R_{\rm b,b}>R_{\rm b,c}$ is left.
That is to say, when the bunch reaches the photosphere, it has already passed the saturation radius.
Under this case, the photosphere radius is given by \citep{2000ApJ...530..292M,2018pgrb.book.....Z}
\begin{eqnarray}\label{eq5}
R_{\rm ph,b}=\frac{\sigma_{\rm T}L_{\rm b}Y}{8\pi \eta_{\rm b}^{3}m_{\rm p}c^{3}}\simeq 2.2\times 10^{11}{\rm cm}\left (\frac{L_{\rm b}}{10^{49}\rm erg\; s^{-1}} \right ) \left ( \frac{\eta_{\rm b}}{30} \right )^{-3}Y,
\end{eqnarray}
where $\sigma_{\rm T}$ is the electron Thomson cross section, $L_{\rm b}$ is the isotropic luminosity of the bunch,
$Y$ is the pair multiplicity parameter, and $m_{\rm p}$ is the proton mass.
As a comparison, the photosphere radius of the jet is
\begin{eqnarray}\label{eq6}
R_{\rm ph,j}=\frac{\sigma_{\rm T}L_{\rm j}Y}{8\pi \eta_{\rm j}^{3}m_{\rm p}c^{3}}\simeq 2.2\times 10^{10}{\rm cm}\left (\frac{L_{\rm j}}{10^{51}\rm erg\; s^{-1}} \right ) \left ( \frac{\eta_{\rm j}}{300} \right )^{-3}Y,
\end{eqnarray}
where $L_{\rm j}$ is the jet power, and $\eta_{j}$ is the dimensionless entropy of the jet.
Thus, it is unnecessary to overly worry about the shielding of the jet to the thermal radiation from the bunch.
However, we should consider the cover of the jet prompt emission to the thermal radiation from the bunch.

According to the normal GRB scenario, the prompt emission is emitted at the internal shock radius which is given by \citep{1994ApJ...430L..93R}
\begin{eqnarray}\label{eq7}
R_{\rm is}\simeq 2\eta_{\rm j}^{2} c \Delta t=6\times 10^{11}{\rm cm}\left ( \frac{\eta_{\rm j}}{100} \right )^{2}\frac{\Delta t}{1\rm ms},
\end{eqnarray}
where $\Delta t$ is the temporal gap between ejecting the two shells of the jet.
When the tail of the jet reaches the internal shock radius, the bunch has arrived at
\begin{eqnarray}
r\sim \beta_{\rm b}c\cdot \left (\frac{R_{\rm is}}{\beta_{\rm j}c }\right )=\frac{\beta_{\rm b}}{\beta_{\rm j}}R_{\rm is},
\end{eqnarray}
where $\beta_{\rm b}=v_{\rm b}/c\approx 1-1/(2\eta_{\rm b}^{2})$, and $\beta_{\rm j}\approx1-1/(2\eta_{\rm j}^{2})$.
Correspondingly, if the termination of the prompt emission is set as the tail of the jet passing the internal shock radius, we have the following three situations.
\begin{itemize}
  \item[I)]
If $r<R_{\rm ph,b}$, the thermal radiation from the bunch will be never covered by the prompt emission.

 \item[II)] When $r>R_{\rm ph,b}$, comparing with the prompt emission, the thermal radiation from the bunch will be emitted earlier at $R_{\rm ph,b}$.
At this time, the tail of the jet has arrived at $\sim R_{\rm ph,b}\beta_{\rm j}/\beta_{\rm b}$.
Therefore, there will be a catch-up problem. In the lab frame, the jet moves with $\beta_{\rm j}$ and leaves $R_{\rm ph,b}$ first.
After a time interval $\delta t$, the thermal radiation leaves $R_{\rm ph,b}$ and begins to pursue the jet (the jet has arrived at $\sim R_{\rm ph,b}\beta_{\rm j}/\beta_{\rm b}$ at this time).
If the thermal radiation from the bunch is observed after the termination of the prompt emission,
the travel distance which makes the thermal radiation from the bunch
catch up with the jet should satisfy (i.e., when the thermal photon arrives at the internal shock radius, it has yet catch up with the jet)
\begin{eqnarray}
2\eta_{\rm j}^{2}c\delta t> R_{\rm is}-R_{\rm ph,b},
\end{eqnarray}
where
\begin{eqnarray}
\delta t&\sim& \left ( \frac{R_{\rm ph,b}}{\beta_{\rm b}c}\beta_{\rm j}c-R_{\rm ph,b} \right )/(\beta_{\rm j}c)\nonumber\\
&\approx& \frac{R_{\rm ph,b}}{\beta_{\rm j}c }\left ( \frac{\beta_{\rm j}}{\beta_{\rm b}}-1\right ).
\end{eqnarray}
In this situation, a small value of $R_{\rm is}$ is needed,
for example, $R_{\rm is}<\sim 100 R_{\rm ph,b}$ by taking $\eta_{\rm j}=100$ and $\eta_{\rm b}=10$.

 \item[III)] Alternatively, when $r>R_{\rm ph,b}$, if the thermal radiation is detected in front of the prompt emission,
there should be
\begin{eqnarray}
2\eta_{\rm j}^{2}c\delta t'< R_{\rm is}-R_{\rm ph,b},
\end{eqnarray}
where
\begin{eqnarray}
\delta t'\sim \left [R_{\rm ph,b}\left ( \frac{\beta_{\rm j}}{\beta_{\rm b}}-1 \right ) +\beta_{\rm j} ct_{\rm 90}\right ]/(\beta_{\rm j}c).
\end{eqnarray}
Similarly, we have $R_{\rm is}>\sim 100R_{\rm ph,b}+2\times 10^{4}ct_{90}$ with the same parameters as that of the situation (II).
\end{itemize}

Once the above three cases are satisfied, a black-body radiation with luminosity \citep{2000ApJ...530..292M,2018pgrb.book.....Z}
\begin{eqnarray}
L_{\rm ph,b}&=&L_{\rm b}\left ( \frac{R_{\rm ph,b}}{R_{\rm b,c}} \right )^{-2/3}\nonumber\\
&=&3.2\times 10^{46}{\rm erg\;s^{-1}}\left ( \frac{L_{\rm b}}{10^{49}\rm erg\;s^{-1}} \right )^{7/12}\left ( \frac{R_{\rm ns}}{1.2\times 10^{6}\rm cm} \right )^{-5/6} \left ( \frac{\eta_{\rm b}}{30} \right )^{8/3}Y^{-2/3},
\end{eqnarray}
temperature
\begin{eqnarray}
T_{\rm obs}&=&\left ( \frac{L_{\rm b}}{4\pi R_{\rm ns}^{2}g_{0}\sigma_{\rm B} } \right )^{-1/4}\left ( \frac{R_{\rm ph,b}}{R_{b,c}} \right )^{-2/3}\nonumber\\
&=&31.3{\rm keV}\left ( \frac{L_{\rm b}}{10^{49}\rm erg\;s^{-1}} \right )^{-5/12}\left ( \frac{R_{\ast}}{1.2\times 10^{6}\rm cm} \right )^{-5/6} \left ( \frac{\eta_{\rm b}}{30} \right )^{11/3}Y^{-2/3}
\end{eqnarray}
and duration being several seconds will be emitted, where $g_{0}$ is the effective degree of freedom of the equipartition theorem, and $\sigma_{\rm B}$ is the Stefan-Boltzmann constant.
Note that,in situation (I) and situation (II), the prompt emission should be followed by the thermal radiation, and in situation (III), the thermal radiation should be followed by the prompt emission.
If an on-axis sGRB with the similar distance as that of GRB 170817A is detected in the future, the discussion shown in this section may can be tested.

\section{Summary and discussion}\label{sec4}

In this paper, we suggest a scenario to eliminate the conflictions between observations and models of GRB 170817A
under the assumption that the remnant of the binary NS merger is an NS (last for, at least, a large fraction of $1\rm s$).
We argue that a fraction of the time delay of $\sim 1.74\rm s$ between GW170817 and GRB 170817A can be resulted by
the absorption of high-energy neutrinos emitted by the central NS and delayed emergence of effective viscous in the disk.
If our scenario is true, when the central object of an on-axis GRB with the similar distance and duration as that of GRB 170817A is an NS,
a weak thermal pulse should be detected before or after the main gamma-ray pulse.
Besides,
our suggestion is hopefully to be tested when the merger remnant is a black hole as long as the delayed jet injection is mainly induced by the delayed disk formation
since, when the merger remnant is a black hole, there should be an evident decrease of the delay.
Once the idea presented here is tested, the time of the delayed jet injection is meaningful to investigate the cooling of nascent NSs, as well as the constraint of the equation of state of NSs.

It is worth noting that, although, we consider a bunch with the similar cross sectional area (CSA) as that of the jet, the CSAs of the two may be different.
As long as the thermal bubble shrinks to satisfy equation (\ref{eq2}), the bunch can break out from the ejecta.
It is no reason to demand that the bunch is exactly shaped by the channel left by the jet.
Therefore, the CSA of the bunch may be larger than the CSA of the jet.
This could be another mechanism to form sheathes of sGRB jets (then the jet-bunch scenario could be a replacement of the jet-cocoon scenario for explaining the thermal tail of GRB 170817A).

\section{Acknowledgement}
We thank the anonymous referee for very useful comments
that have allowed us to improve our paper.
We would like to thank Drs. Liang Li and Chengjun Xia for useful discussion.
This work is supported by the National SKA Program of China (2020SKA0120100), research projects of Henan science and technology committee (212300410378)
and the NSFC grants (U1938116).



\section{Data Availability}
no new data generated.

\end{document}